\documentclass{aa}  

\usepackage{graphicx}
\usepackage{txfonts}
\usepackage{float}
\usepackage{subfigure}
\graphicspath{ {/home/ummi/Gareq/TechNote/allStars/} }

\usepackage{soul}
\usepackage{xcolor}
\sethlcolor{green} 
\begin{document}

\title{Differential Astrometry with Gaia:}

   \subtitle{Investigating relativistic light deflection close to Jupiter}

   \author{U.~Abbas
          \inst{1}\fnmsep\thanks{\email{ummi.abbas@inaf.it}}
          \and B.~Bucciarelli\inst{1}\and M.G.~Lattanzi\inst{1}\and M.~Crosta\inst{1}
          \and R. Morbidelli\inst{1}\and D.~Busonero \inst{1}
          \and L. Bramante\inst{2}\and R. Messineo\inst{2}
          }

   \institute{INAF - Osservatorio Astrofisico di Torino, Via Osservatorio 20,
              I-10025 Pino Torinese, Italy \\
              \and
              ALTEC S.p.a, Corso Marche, 79, 10146 Torino, Italy}
\titlerunning{Differential Astrometric analysis of the GAREQ experiment}
\authorrunning{U.~Abbas et al.}

   \date{Received May 9, 2022; accepted June 14, 2022}

 
  \abstract
   {}
   {In this paper we develop a differential astrometric framework that is appropriate 
   for a scanning space satellite such as Gaia. We apply it to the first of the GAREQ fields - the Gaia   Relativistic 
   Experiment on Jupiter’s quadrupole - which is the fruit of dedicated efforts within the Gaia project
   focused on measuring the relativistic deflection of light close to Jupiter's limb. This provides 
   a preliminary assessment of: a) the observability of the relativistic deflection of light close to Jupiter, 
   and, b) Gaia's astrometric capabilities under extremely difficult conditions such as those around a bright extended object.
   }
   {Inputs to our differential astrometric model are the charge-coupled device (CCD) transit times as measured by Intermediate
   Data Update (IDU) system, transformed
   to field angles via Astrometric Global Iterative Solution (AGIS) geometric calibrations, 
   and the commanded/nominal spacecraft attitude.
   Actual attitude rates, including medium and high-frequency effects, are estimated from successive 
   CCD pair observations and used to transfer the field angles onto intermediate tangent planes, finally
   anchored to a common reference frame by fitting a 6-parameter model to a set of suitable 
   reference stars.
	The best-fit parameters provide the target star’s deflection as a time-varying systematic effect.
	To illustrate the model we analyze Gaia astrometric 
   	measurements after their calibration through the latest cyclic EDR3/DR3 processing 
   	of the GAREQ event of February 2017. 
 	We use observations of the closest bright target star successfully observed several times by Gaia 
	in close proximity to Jupiter and surrounding 
	reference stars brighter than G<13 mag in transits leading up to the time of 
	closest approach and on subsequent transits.
	}
   {The relativistic deflection signal is detected with a S/N of 50 at closest approach by 
   the target star.
   This signal is the combined effect due to Jupiter
   and the Sun, mainly dominated by Jupiter's monopole, demonstrating 
	Gaia's scientific performance under extreme observational conditions. 
     It is an unprecedented detection for the following reasons: 
	a) closest ever to Jupiter’s limb ($\sim$7") in the optical, and,
      	b) highest S/N at any wavelength.    
     Finally, this work sets the stage for investigations into disentangling
     the relativistic quadrupole deflection due to Jupiter with future Gaia astrometric measurements.
     }
   {}


   \maketitle
%

\section{Introduction}

The Gaia space satellite is currently taking measurements of 1.8 billion sources 
brighter than $G<$ 21 mag \citep{brown2021} and is set to provide their 5-parameter global
astrometric solutions to unprecedented precision continuously improving with each data release.
The global measurement principle of Gaia for such a
large number of stars has led to the need for a sophisticated machinery that entails the Astrometric Global
Iterative Solution (AGIS) and involving a block-iterative least-squares solution of the astrometric, attitude,
calibration and global parameters in a cyclic manner until convergence \citep{lindegren2012}.
If one is interested only in a small field of sources, then differential astrometry can be used to
model transient astrometric events and/or unmodeled systematic effects.

In the context of differential astrometry one requires the establishment of a small field inertial
reference frame prior to studying such physical effects, e.g. in cluster membership studies, detecting the
reflex motion of the host star due to its planets, obtaining trigonometric parallaxes, measuring the relativistic
deflection of light due to giant planets to name a few (see \citealt{abbas2017} and references therein).
For space-based observations and in the absence of Earth's atmosphere many problems are avoided, nonetheless there are
effects due to the: a) light aberration of the order of $\sim$20 arcseconds to first order and few mas to second
order \citep{lattanzi1993}, b) gravitational deflections due to various moments of giant planets and the Sun
\citep{crosta2006}, c) parallaxes and proper motions of stars \citep{lattanzi1993, vanAltena2013}, and,
d) changes in the geometric instrument model due to imperfections in the optics, charge-coupled device (CCD) structure or positioning, and, 
thermal variations in the instrument \citep{lindegren2012}.

According to \cite{crosta2006} the motion of a giant planet such as Jupiter along its orbit causes favorable scenarios where Jupiter,
as seen by Gaia, lies in close proximity (i.e. within ten arcseconds from its limb) to relatively bright stars.
Under such conditions the GAia Relativistic Experiment on Quadrupole (GAREQ) light deflection (at that epoch known as 
GAREX - Gaia Relativistic Experiment) was designed in order to 
study the deflection of 'grazing' starlight due to Jupiter's monopole and quadrupole moment.
Firstly, a galactic model was used and secondarily a restricted field of stars was chosen from a real catalog (GSC2; 
\citealt{lasker2008}) in order to 
simulate the detection of the effect due to a bright star passing close to Jupiter's limb \citep{crosta2008a}.
Such situations generally arise as Gaia scans the sky close to the galactic plane 
and depend on the initial scanning law parameters \citep{deBruijne2010}. 
The follow-up optimization campaign carried out by the dedicated RElativistic Modeling
And Testing (REMAT) working group within the Data Processing Analysis Consortium (DPAC) of Gaia
with a further fine-tuning of the spin phase led to the predicted favorable
configuration of three stars with $G<$ 15.75 mag close to Jupiter's limb for February 2017 \citep{klioner2014a, 
klioner2014b, abbas2014}.
The high-cadence observations obtained by Gaia in February 2017 were highly successful with some barely a few arcseconds
from Jupiter's limb, despite the stray light from the extremely bright Jupiter disk that strongly affects
observations within a couple of arcseconds from its limb even capable of washing out entirely any
given observation. 

The first detection of relativistic deflection by a body other than the Sun, i.e. Jupiter, was seen  
in the near-occultation event of 21 March 1988 with the Very Long Baseline Interferometry (VLBI) measurements, 
when the ray path of the radio source P0201 + 113 came within 200 arcsec of Jupiter \citep{treuhaft1991}.
In their case, the expected gravitational bend of the ray path according to general relativity, averaged over the
experiment duration, was approximately 300 $\mu$arcsec with measurement accuracies of the order of 160 $\mu$arcsec.
Other attempts in the optical with the Hubble Space Telescope were made \citep{whipple1996}, but their results were never published.

In the spirit of \cite{abbas2017} where the authors made a first attempt at using simulated Gaia
measurements to look at the short term stability of a Differential Astrometric reference frame, we
will adopt a similar approach and study the stability of the small field reference frame
and the relativistic deflection of light during the GAREQ event of February 2017.
In our differential analysis we focus on the closest brightest star with $G=12.78$ mag 
(hereafter to be called the {\it target star}) that was
seen a total of 26 times over a 2 month period starting at the beginning of 2017, out of which we use
25 transits over a short time interval of roughly 3 days surrounding the observation at closest approach.

The paper is divided as follows: Sec.~\ref{sect:alldiff} presents the all-differential astrometric framework, 
Sec.~\ref{sect:observations} shows the details of the Gaia observations that are subsequently used in the model and
their cleaning, Sec.~\ref{sect:effects} describes the physical and instrumental effects on the
observations, 
Sec.~\ref{sect:results} illustrates the relativistic deflection of light during the GAREQ event, Sec.~\ref{sect:error_budget}
discusses the error budget due to the reference stars,
finally wrapping up with the conclusions in Sec.~\ref{sect:conclusions}.

\section{The All-Differential Astrometric Model}\label{sect:alldiff}

The fundamental ingredients to the differential astrometric model are the Satellite Reference System (SRS) field angles, $\eta'_{ij}(t_{ij})$ and $\zeta'_{ij}(t_{ij})$,
of the $j$th star in the $i$th frame\footnote{A frame is given by the observations of stars over one transit where 
these observations are obtained with the AF1-9 CCD column fiducial lines; $t_{ij}$ obtained with any given CCD column
lies within $\Delta t = $40s from the {\it target star} observing time for the same column. Then the AL/AC rate is used to
transform $t_{ij}$ for that transit to the {\it target stars} observing time at AF5.}, in the along (AL) and across (AC) 
scanning directions respectively,
with the best available geometric calibration at each OBMT\footnote{The on-board mission timeline (OBMT) is conveniently used
to label on-board events; it is expressed as the number of nominal revolutions of exactly 6 hours on-board time from
an arbitrary origin.} observation time $t_{ij}$ (a description of how these are obtained is given in the next Section).
The {\it target star} as measured in the 
$i$th frame is denoted by $\eta^{'}_{i0}(t_{i0})$ and $\zeta^{'}_{i0}(t_{i0})$.
 
All the field angles at different times within a given frame need to be referred to the observing time unique to that frame 
(for convenience, we choose this to be the observing time $t_{i0}$ of the {\it target star} for that frame). 
This is done 
by applying a correction to the field angles which depends on their rate of change, i.e.,  
\begin{align}
\eta'_{ij}(t_{i0})  &= \eta'_{ij}(t_{ij})+\sum_{t_{i0}}^{t_{ij}}\dot \eta[\eta(t),
\zeta(t),t]\Delta t \nonumber \\
\zeta'_{ij}(t_{i0}) &= \zeta'_{ij}(t_{ij})+\sum_{t_{i0}}^{t_{ij}}\dot\zeta[\eta(t),\zeta(t),t]\Delta t,
\label{eq:rateChange}
\end{align}
with $\Delta t$ chosen to be 1sec.

The field-angle rates are expressed by the analytical formulae (see \citealt{lindegren2018}, Eq.~4):
\begin{align}
  \dot\eta[\eta(t),\zeta(t),t] &= -\omega_z + [\omega_x \cos\varphi + \omega_y \sin\varphi] \tan\zeta(t) \nonumber \\
  \dot\zeta[\eta(t),\zeta(t),t] &= - \omega_x \sin\varphi + \omega_y \cos\varphi
\label{eq:etaDot}
\end{align}
where $\omega_x, \omega_y$ and $\omega_z$ are the components of the instantaneous inertial angular velocity of Gaia, and
$\varphi = \eta \pm \gamma/2$, where the plus or minus sign is used for preceding or following FOV respectively and $\gamma$
being the Basic Angle\footnote{the angle between Gaia's two fields of view.}. 

Estimates of the inertial angular rate along the SRS axes x, y and z are obtained from pairs of successive 
CCD observations in the astrometric
field (AF) of the same source with corresponding successive observation times
(further detailed in Sec.~3.2 of \citealt{lindegren2018}). In short, this 
involves a two-step procedure whereby the measured rates are first reduced to the center of the field (where $\eta$ = $\zeta$ = 0), 
and transformed to the inertial angular velocity components in a second step.
At the field center: $\dot\zeta^0 = \pm\omega_x \sin\gamma/2 + \omega_y\cos\gamma/2$ which when subtracted from
$\dot\zeta$ gives:
\begin{equation}
	\dot\zeta^0 \simeq \dot\zeta + (\omega_x \cos\varphi + \omega_y \sin\varphi)\sin\eta	
\label{eq:zeta0Dot}
\end{equation}
after neglecting second order terms in $\eta^2$. 
$\dot\zeta$ are essentially the raw rates that are calculated using all suitable AC pair observations and are simply
the differences in the AC field angle between pairs of successive CCD observations divided by their corresponding time 
differences. Using Eq.~\ref{eq:zeta0Dot} these are then corrected for the field rotation by means of the commanded inertial angular
rates $\omega_x$ and $\omega_y$.
For each FOV separately, the values $\dot\zeta_{P/F}^0$ are sorted by time and the running
n-point medians are computed to obtain arrays of $\dot\zeta_{P/F}^0$ from which the required value can be interpolated.

$\omega_x, \omega_y$ can then be calculated as follows:
\begin{equation}
  \omega_x = -\frac{\dot\zeta_P^0 - \dot\zeta_F^0}{2 \sin\frac{\gamma}{2} }, \qquad
  \omega_y = +\frac{\dot\zeta_P^0 + \dot\zeta_F^0}{2 \cos\frac{\gamma}{2} }
\label{eq:omXY}
\end{equation}
where the required value of $\dot\zeta_{P/F}^0$ is interpolated to the time of the rate estimate.

The same procedure is adopted for $\omega_z$ using 
\begin{equation}
	\dot\eta^0 = \dot\eta - (\omega_x \cos\varphi + \omega_y \sin\varphi)\tan\zeta	
\label{eq:eta0Dot}
\end{equation}
where $\dot\eta^0 = -\omega_z$ has been replaced in Eq.~\ref{eq:etaDot}a and is calculated using the AL raw rates ($\dot\eta$) 
and the improved rates $\omega_x, \omega_y$ from Eq.~\ref{eq:omXY} above.
The rate measurements, $\dot\eta^0$, are binned by time, using a bin size
of 0.5 s, and the median values are calculated in each bin providing an accurate time-series representation of $\omega_z(t)$. This bin size is for maximum time resolution and just large enough to allow for a reliable (robust) estimate per bin requiring a minimum of some 10 to 20 values per bin.

\begin{figure*}
\centering
    \includegraphics[width=\hsize]{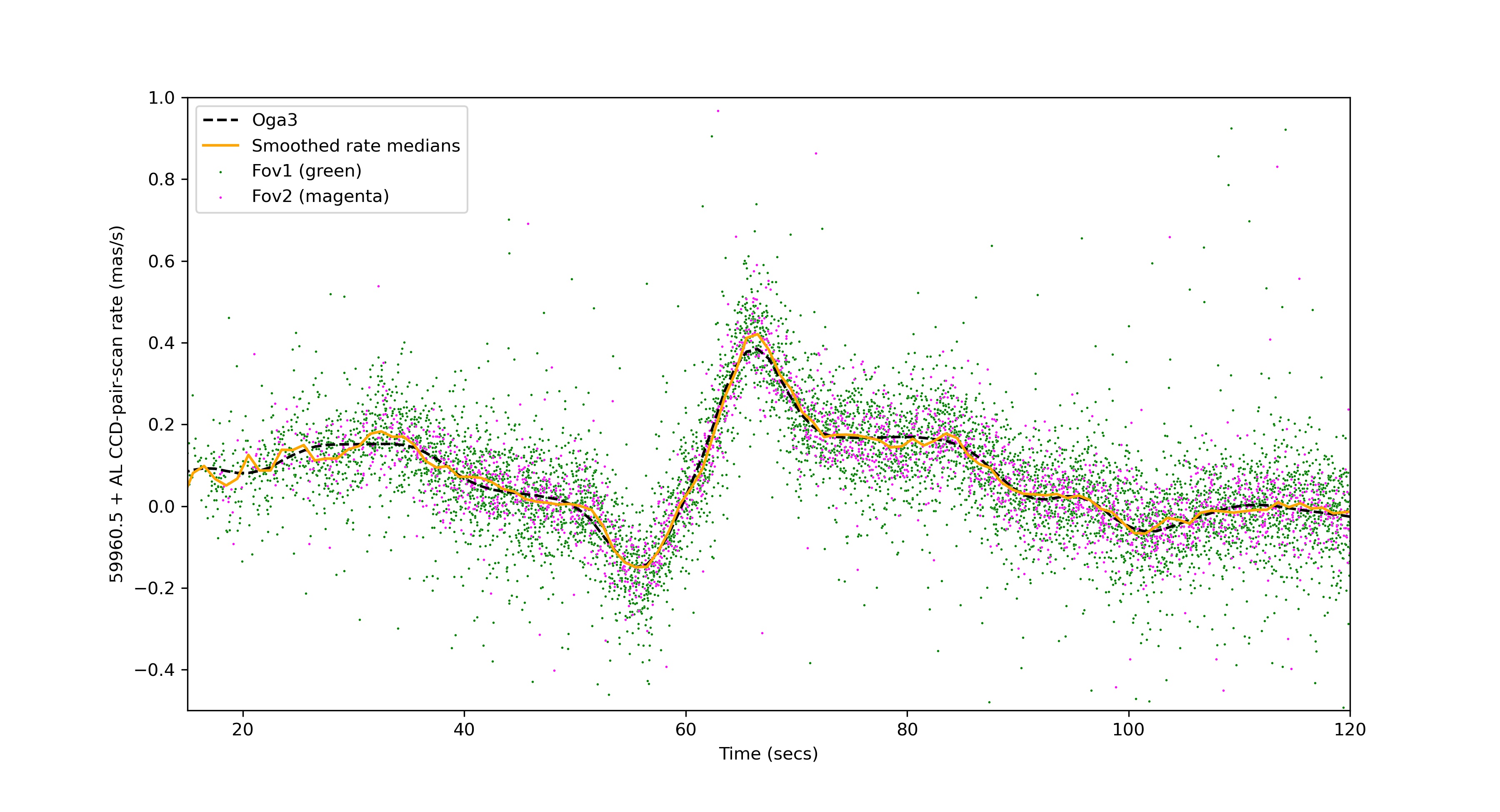}
    \caption{CCD pair scan AL rates for transit no. 1. The green and magenta points show
	the pair scan rates for FOV1 and FOV2 respectively, while the rate medians obtained in bins of 
	0.5 sec for these measurements is shown by the orange line that connects each point separated by 1sec. 
	The black dashed line shows the AL rate as obtained with OGA3.2 (including the corrective attitude),
	at the same times as for the orange line.
      }
\label{fig:pairRates}
\end{figure*}

An example of the effectiveness of this procedure is shown as the orange line in Fig.~\ref{fig:pairRates} for the first transit of the {\it target star} during the GAREQ event of Feb 2017(for more examples see Appendix A of \citealt{abbas2021}).
This can be compared to the third on-ground attitude determination (OGA3.2) AL rate and the good agreement demonstrates 
the accuracy in obtaining the AL rates in this manner.

Once the field angles of each star have been transformed to their proper frame time, we estimate and remove a priori the effects due to Gaia EDR3 proper motions (PMs) and parallaxes 
(Sec.~\ref{sect:apriori}). 
These field angles are then transformed by a rotation and then a gnomonic transformation (Sec.~\ref{sect:gnomonic}). 
The least-squares fitting procedure 
then involves the Taylor series expansion of the transformed field coordinates (Sec.~\ref{sect:taylor}) 
and finally a linear plate transformation onto the reference frame (Sec.~\ref{model}).

\subsection{A priori removal of parallaxes and proper motions}\label{sect:apriori}
Here we show the procedure 
of removing a priori the parallaxes and PMs from Gaia EDR3. 
This step is necessary as unmodeled astrometric parameters add systematic effects with 
unmodeled PMs having a stronger effect than unmodeled parallaxes, 
$\sim$23 $\mu$as (AL) and 69 $\mu$as (AC) due to unmodeled PMs versus the $\sim$5 $\mu$as (AL) and
17 $\mu$as (AC) due to unmodeled parallaxes \citep{abbas2019}. 
These quantities are given in equatorial coordinates that then need to be converted to the Satellite 
Reference System (SRS) giving the projected along and across scan values. 
This transformation for PMs is completely determined by the position angle of the scan (computed from 
the nominal attitude)
and is given by:
\begin{align} \label{eqn:transform_scan}
    \mu_\eta &= \mu_\alpha \sin \theta + \mu_\delta \cos \theta
  \nonumber \\
    \mu_\zeta &= - \mu_\alpha \cos \theta + \mu_\delta \sin \theta
\end{align}
When treating the parallaxes, $\mu_\alpha$ and $\mu_\delta$ can be replaced with the parallax
factors $f_\alpha$ and $f_\delta$ respectively, calculated using star catalog values and Gaia ephemerides, 
to obtain the transformed parallax factors along 
and across the scanning direction.
The `corrected' field angles are then given as:
\begin{align} \label{eqn:apriori}
    \eta'_{ij}(t_{i0}) &\equiv \eta'_{ij}(t_{i0}) - \mu_\eta(t_{i0}-T) - f_\eta \pi
  \nonumber \\ 
    \zeta'_{ij}(t_{i0}) &\equiv \zeta'_{ij}(t_{i0}) - \mu_\zeta(t_{i0}-T) - f_\zeta \pi
\end{align}
where $\eta'_{ij}(t_{i0})$ and $\zeta'_{ij}(t_{i0})$ at time $t_{i0}$ on the RHS are calculated as given 
at the very beginning of this Section, and $T=t_{10}$ is the observing time of the {\it target star} on the 1st frame (arbitrarily chosen).

\subsection{Pre-rotation and Gnomonic transformation}\label{sect:gnomonic}

Before trying a global adjustment of all the frames using the principles of tangent-plane astrometry, field angle measurements must be rectified via a gnomonic transformation. 
In order to minimize the differential effect of the so-called $\it tilt$ terms, which are second-order quantities arising from a misalignment of the nominal vs. true 
position of the telescope's optical axis, all the reference stars, including the target, are rotated in such a way that the position of the {\it target star} becomes aligned with the (1,0,0) vector defining the origin of the $\eta$, $\zeta$ coordinates.

Then these field angles are converted to standard coordinates in the tangent plane with tangent point (0,0)
using gnomonic transformations as follows:
\begin{equation}\label{eqn:gnomonic}
  X^{'}_{ij} = \tan(\eta^{'}_{ij}), \qquad 
  Y^{'}_{ij} = \frac{\tan(\zeta^{'}_{ij})}
  {\cos(\eta^{'}_{ij})}
\end{equation}
for the $i$th frame and the $j$th star at the time $t_{i0}$ of the {\it target star} for that frame (the time index is 
dropped from now on as all the times hereafter refer to this time).

\subsection{Taylor series expansion}\label{sect:taylor}
We linearize the expressions from Eq.~\ref{eqn:gnomonic} and perform a Taylor series expansion of the reference star coordinates in the tangent plane 
from above around the calibrated position of the star. 
The expansion to first order can be written as:
\begin{align} \label{eqn:taylor}
  X^{'}_{ij} &= X'_{ij}\Bigr|_{\substack{\eta'_{ij} \zeta'_{ij}}} 
  + \Delta\eta_c \frac{\partial X'_{ij}}{\partial \eta'_{ij}} + 
        \Delta\zeta_c \frac{\partial X'_{ij}}{\partial \zeta'_{ij}}
  \nonumber \\  
  Y^{'}_{ij} &= Y'_{ij}\Bigr|_{\substack{\eta'_{ij} \zeta'_{ij}}} + \Delta\eta_c \frac{\partial Y'_{ij}}{\partial \eta'_{ij}} + 
        \Delta\zeta_c \frac{\partial Y'_{ij}}{\partial \zeta'_{ij}}
\end{align}

where the terms, $\Delta\eta_c$ and $\Delta\zeta_c$, are treated as unknowns that are simply 
small corrections to the calibrated field angles whereas
the partial derivatives are evaluated at the calibrated position. Upon closer investigation we 
found that the estimated values of the calibration unknowns are rather insignificant (at the sub-$\mu$as level) and hence these were eliminated thereby leaving us only with the first term.

\subsection{The Differential Astrometric model}\label{model}
The positions of the stars in each frame, at time $t_{i0}$, are therefore adjusted to its reference 
frame value through a simple plate transformation given as:
\begin{align} \label{eqn:full_linPM}
  X^{'}_{0j} &= a_i X^{'}_{ij} + b_i Y^{'}_{ij} + c_i 
  \nonumber \\
 Y^{'}_{0j} &= d_i X^{'}_{ij} + e_i Y^{'}_{ij} + f_i 
\end{align}

where $X^{'}_{ij}$ and $Y^{'}_{ij}$ are the star
coordinates in the tangent plane, with $X^{'}_{0j}$ and $Y^{'}_{0j}$ being the coordinates of the $j$th star on the
reference frame and the linear plate constants are given by $a_i$, $b_i$, $c_i$, $d_i$, $e_i$ and $f_i$.

The `transformed' field coordinates, $\eta'_{0j}$ and $\zeta'_{0j}$ at times $t_{i0}$, 
are then obtained through the inverse gnomonic transformation (from Eq. \ref{eqn:gnomonic}).

The traditional {\it all-differential} approach that has been published in \cite{abbas2019} 
typically defines a `plate' as all stars clocked by a single fiducial CCD line, i.e. each FOV transit consists of 9
plates from AF1-9. The reason being that a linear plate adjustment accounts for rotations and 
translations between the various plates. 
If one uses the AL/AC rate to account for the rotation over one transit, 
i.e. for AF1-9, then the transit itself can be treated as a plate. 

We have 90 unknown linear plate parameters (by treating each transit as a plate) 
to be estimated
and a total of 3364 AF1-9 observations (after removing flagged observations) from the 15 transits in Table~\ref{table:1}
(following Section).
The software package GAUSSFit \citep{jefferys1988}
is used to solve this set of equations through a generalized least-squares procedure weighted according to the input errors (obtained from the standard errors in the
calculation of the AL/AC centroid in AF1-9 as estimated by the centroiding algorithm of the Gaia processing pipeline). 
GAUSSfit uses an iterative process whose convergence is controlled by the input tolerance.
Moreover, we perform an external iteration loop to reject outliers until a low-skewness (<0.1) gaussian distribution of the residuals is achieved.
On average, anywhere from 0.04-1.5\% (depending on the field of interest) of the observations  are deemed problematic and not used in the final fit.
The estimated plate/frame parameters ($a_i$ through $f_i$) allow to transport the calibrated observations ($X^{'}_{ij}$, $Y^{'}_{ij}$) to a common reference frame. 
If all other effects are properly modeled the distribution of residuals should show the unmodeled Sun and Jupiter 
monopole (plus Jupiter quadrupole) deflection terms of the reference stars that have not been subtracted out.
As can be expected, it is found that $a_i$ and $e_i$ are almost unity,
whereas $b_i$ = $-d_i$  and together they give the rotation and orientation. 
The offsets $c_i$ and $f_i$ give the offset of the common system.
One can try to further reduce the number of unknown plate/frame parameters by imposing a pure rotation 
and scale, we found that this gives similar results.

\section{Handling of the observations}\label{sect:observations}

To illustrate the differential model developed we will use the astrometric measurements 
taken by the Astrometric Field (AF) CCD detectors in the Gaia focal plane (see Fig.~3 in \citealt{lindegren2021}). 
These observations are taken in Time-Delayed 
Integration (TDI) mode so as to allow for the accumulation of charge as the images move
across the CCDs due to the spinning motion of Gaia \citep{brown2016}. 
The fundamental astrometric observational quantity is given
by the OBMT time corresponding to the passage of a stellar image centroid through the
fiducial line of a CCD (typically halfway
between the first and last TDI line used in the integration for the CCD) encapsulated in the elementary astrometric 
observations processed by the intermediate data update (IDU) system. 
For each transit 9 observations were obtained from the AF 1 to 9 columns.
The input AF observations are given in the window reference system (WRS) that contains information of each detection,
more specifically the CCD, pixel, gate and the Field of View (FOV) of Gaia (see \citealt{fabricius2016} for full details).
This information  
is used, along with the instrumental calibrations from the cyclic processing by AGIS, to convert the star's observing time into its position in the SRS system, i.e. the calibrated field angles in the AL and AC scanning directions \citep{lindegren2021} that are ultimately useful for this analysis. 

The $\eta$(AL) and $\zeta$(AC) field angles mentioned above are 
essentially the spherical coordinates
on the sky relative to a reference direction in each FOV \citep{lindegren2012}.
These represent the fundamental input to the model along with an ulterior adjustment using the commanded/nominal attitude
(the reason for this operation is further described in Sec.~\ref{sect:alldiff}).

\begin{table*}[h]
  \centering
  \begin{tabular}{cccccc}
\hline
TransitId & observed OBMT[long] & UTC & FOV & CCD row & b [Rjup] \\
 \hline \hline
1 & 104799283188665306 & 2017-02-22T19:08:02.862  & 1 & 7 & 4.29
\\ 
2 & 104805682340958154 & 2017-02-22T20:54:42.015  & 2 & 7 & 3.61
\\
3 & 104827296775979656 & 2017-02-23T02:54:56.450  & 2 & 6 & 1.35
\\
4 & 104864126189227205 & 2017-02-23T13:08:45.862  & 1 & 2 & 2.85
\\
5 & 104870525213255710 & 2017-02-23T14:55:24.887  & 2 & 3 & 3.55
\\
6 & 104885740271535278 & 2017-02-23T19:08:59.945  & 1 & 2 & 5.24
\\
7 & 104892139269141833 & 2017-02-23T20:55:38.943  & 2 & 3 & 5.96
\\
8 & 104907354273083680 & 2017-02-24T01:09:13.947  & 1 & 1 & 7.67
\\
9 & 104913753251905106 & 2017-02-24T02:55:52.925  & 2 & 2 & 8.40
\\
10 & 104928968221338313 & 2017-02-24T07:09:27.895  & 1 & 1 & 10.14
\\
11 & 104935367189558312 & 2017-02-24T08:56:06.863  & 2 & 2 & 10.88
\\
12 & 104950582144816169 & 2017-02-24T13:09:41.818  & 1 & 1 & 12.64
\\
13 & 104956981108577159 & 2017-02-24T14:56:20.782  & 2 & 2 & 13.38
\\
14 & 104972196070729328 & 2017-02-24T19:09:55.744  & 1 & 1 & 15.17
\\
15 & 104978595039079886 & 2017-02-24T20:56:34.713  & 2 & 2 & 15.92
\\ \hline  \\
\end{tabular}
\caption{The list of observed transits for the {\it target star} close to Jupiter. Column 2 shows the 
Sky Mapper observed OBMT from the {\it target star}'s elementary astrometric observation, column 3 gives the corresponding time in UTC obtained 
with the GOST tool, columns 4 and 5 give the FOV and CCD row number respectively of the transit and the last column
is the estimated impact parameter 'b' from Jupiter's center in units of Jupiter radii.}
\label{table:1}
\end{table*}

\begin{figure*}
  \centering
      \includegraphics[width=\hsize,scale=0.6,trim =0cm 0cm 0cm 0cm]{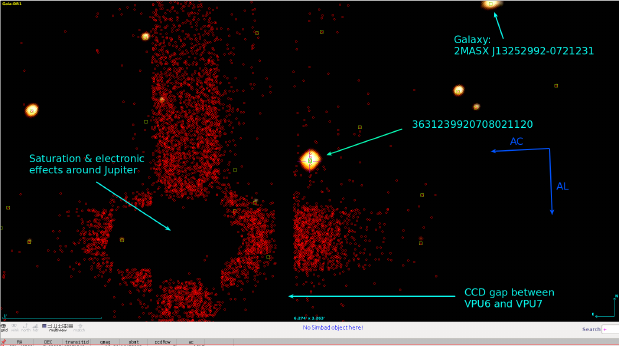}
      \caption{Scene around the {\it target star} successfully detected onboard during the transit on 
	2017-02-22T19:08:12.954 (UTC) at a predicted angular distance of 67.65” from Jupiter's limb. 
	Note the gap between the detectors in CCD rows 6 and 7. Whilst Jupiter was observed in row 6, 
	the {\it target star} happened to be scanned in row 7. 
        Image credit: ESA/Gaia/DPAC/C. Crowley)}
  \label{fig:scene}
\end{figure*}

We remind the reader that Gaia's astrometric instrument is optimized for measurements in the AL 
scanning direction, with 
rectangular pixels that are three times larger in the AC direction \citep{prusti2016}.
This translates into larger uncertainties in the AC observations versus those in the AL direction. 
Furthermore, only stars brighter than $G< $13 mag have two-dimensional windows resulting in accurate AC 
positional information (accurate AL measurements are always available). 
Fainter stars are typically acquired as one-dimensional in the AF due to the on-board removal of 
AC position information by on-chip binning. At the faint end some 'calibration' stars sporadically have two dimensional 
information, otherwise one-dimensional windows always come with the Sky Mapper AC observations providing an approximate position in the AC direction at the pixel level.

\begin{figure*}
  \centering
      \includegraphics[width=\hsize,scale=0.6,trim =0cm 0cm 0cm 0cm]{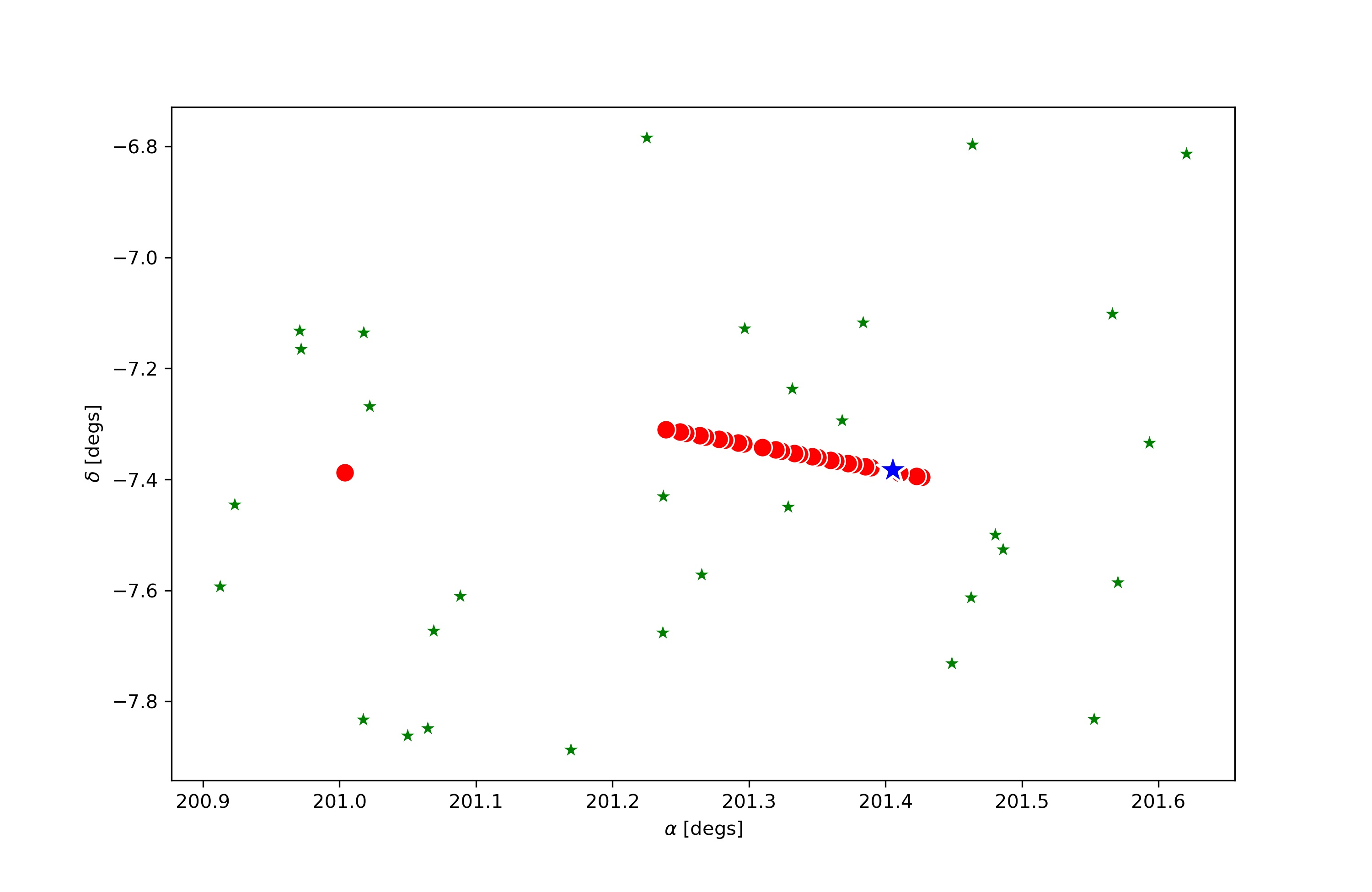}
      \caption{The star field surrounding the brightest star closest to Jupiter's limb in the
		magnitude range of 10 $<$ G $<$ 13 mag within 0.8X1.3 degs on the sky along with
		Jupiter using the INPOP10 ephemeris. The small green star symbols show the stars, the big red
		dots denote Jupiter's position as seen on each transit by Gaia when the {\it target star} was observed
                (the single large red dot on the left part is the Jan transit, and the set of 25 dots in the
                center are the Feb transits), and, the {\it target star} is shown as the large blue star
                at $\alpha$ = 201.41, $\delta$ = -7.38 degs.}
  \label{fig:1}
\end{figure*}

We select stars surrounding the {\it target star} during the GAREQ event of February 2017 based on 
their observation times 
to lie within a chosen interval for any given time of observation of the {\it target star} shown in Table~\ref{table:1}.
This table also gives the estimated distance from Jupiter's center for 15 transits that were used in this study.
There are two transits that were predicted and not
observed by Gaia due to different reasons, 
in one case the observation was contaminated by a Jovian Moon lying almost on top of the star
and in the other case it fell in the gap between CCD rows \citep{crowley2017}. 
Fig.~\ref{fig:scene} is an example of the scene around the {\it target star} successfully detected onboard during the first transit in 
Table~\ref{table:1} and shows the various effects of heavy saturation in the detection record overlay.

Fig.~\ref{fig:1} shows the stars with magnitudes 10$<$G$<$13 surrounding the {\it target star} based on such a selection, 
in this case 40 secs. Superposed on the same field are the 
estimated Jupiter positions (Gaia-centric) from the INPOP10 ephemeris file at the observed OBMTs of
the {\it target star} as seen in January and February 2017, including the 15 transits from Table~\ref{table:1}.
The single large red point in the left half of the figure corresponds to a single transit by Jupiter in January, 
the line of 25 consecutive large red points (several pairs of which are overlapping) in the right hand part instead correspond to the February transits. 
The last 10 transits are used in the 
error-budget analysis, as they allow to study the different systematics produced by Jupiter still in the 
field but further 
away (>300 arcseconds), and is further discussed in Sec.~\ref{sect:error_budget}.
All AF observations were used after properly cleaning them as mentioned in the next paragraph.
The differential analysis was carried out using the stars shown in Fig.~\ref{fig:1}, i.e. bright stars
that have 2d windows and therefore the AL ($\eta$) and AC ($\zeta$) coordinates.

The observations have various flags based on the on-board acquisition and processing results at the 
transit and CCD level. 
In general, we found that it was necessary to clean the observations that had cosmetic issues in the processing 
and with saturated samples removed. At times there were also some due to an abnormal charge injection preceding 
the window in the acquisition. Furthermore, only observations with goodness-of-fit values of the image fit to the 
point/line spread function (PSF/LSF) model of less than 100 were retained in order to eliminate the few extreme outliers.
The {\it target star} of interest had a few similar flags, as well as some due to a non-nominal 
gate\footnote{Sources brighter than G$\sim$12 lead to saturated images mainly due to the sensitivity of 
the astrometric instrument. TDI gates mitigate this effect, these
are special structures on the CCDs that can be activated in order to 
inhibit charge transfer and hence to effectively reduce the integration time for bright sources.} 
in the window and multiple gates, especially when it was observed closest to the limb of Jupiter, i.e. TransitId 3 of Table~\ref{table:1}.

\section{Physical and Instrumental effects}\label{sect:effects}
The measurements are generally affected by astrometric 
effects such as velocity aberration and gravitational light
deflection, and by the proper motions and parallaxes of the
sources which can all be classified as physical effects. They are
also subject to instrumental errors that require accurate
modeling.
We will briefly summarize each of these, for more details please refer 
to Sec.~3 of \cite{abbas2017}.

\subsection{Physical effects}
The aberration is by far the dominant effect and is caused by
the motion of the observer with respect to the barycenter of the
solar system (\citealt{klioner1992}; see also appendix in \citealt{lattanzi1993}).
It is roughly of the order of v/c to first order. For
the speed of Gaia ($\simeq$29.6 km/s ) 
the maximum values
(projected values along the AL direction) are roughly 20".3
to first order, $\sim$2.7 mas to second order, and third order terms
are at the 1 $\mu$as level.

The gravitational deflection of light due to Solar System
objects is another major effect that needs to be taken into
account and depends on the angular separation between the
Solar system body and the given source. In the GAREQ field mainly
Jupiter and the Sun give contributions, where
the dominant deflection effect is that due to Jupiter reaching 16.2~mas at its limb and falling off
as the inverse of the impact parameter (principally due to Jupiter's monopole; the quadrupole 
deflection effect is barely 240$\mu$as at the limb that instead decreases 
as the inverse cube of the impact parameter). The deflection due to the
Sun amounts to $\sim$1.8 mas (with a 0.3-0.6mas effect in AL) as seen by Gaia
for this GAREQ field.
Over timescales of 24 hours the differential aberration is a largely linear effect that
amounts to several mas AL scan for this field (see Fig. 4 in \citealt{abbas2017}), 
whereas the differential gravitational deflection
is non-linear contributing several mas (mainly due to Jupiter's monopole) 
and sub-$\mu$as due to Sun's monopole.

The stars proper motions (PMs) can vary up to several hundreds of mas/yr, 
for this particular field the set of reference stars has Gaia EDR3 PMs varying up to 116 mas/yr 
and with parallaxes up to 9 mas. The effect due to PMs is mostly linear
and can be of the order of tens of $\mu$as over 24 hours.

\subsection{Instrumental effects}\label{sect:calibrations}

The observation lines, given by the fiducial lines mapped
onto the tangent plane, are affected by the geometric instrument
model describing the layout of the CCDs. This includes the
physical geometry of each individual CCD and its configuration 
in the Focal plane assembly; the distortions and aberrations
in the optical system; nominal values of the focal length and
Basic Angle, 
$\gamma$ (see \citealt{lindegren2012, lindegren2016} for extensive details).
These effects are time-dependent and one of three types: large-scale AL 
calibrations occuring on short timescales, small-scale AL calibrations expected to be stable 
possibly over the whole mission duration, and, large-scale AC calibrations 
assumed to be constant on long timescales.

As we are considering observations over a few days, we 
shall only be concerned with the large scale AL and AC calibrations that can 
be taken to be constant to first approximation.
The AL (and AC) large-scale calibration is modeled as a low order polynomial in the across-scan pixel coordinate 
$\rho$ (that varies from 13.5 to 1979.5 across the CCD columns, \citep{lindegren2012}

The effect due to calibrations is highly non-linear at the mas level.
The handling of the calibrations and unmodeled errors during the cyclic processing
by AGIS improves with each cycle (see Fig.~9 in \citealt{lindegren2018} versus Fig.~A.1 in 
\citealt{lindegren2021}) and is significantly better than the calibrations from the daily processing
pipelines.

\newpage
\section{The Feb 2017 GAREQ deflection event}\label{sect:results}

\begin{figure*}
  \centering
      \includegraphics[scale=1.1]{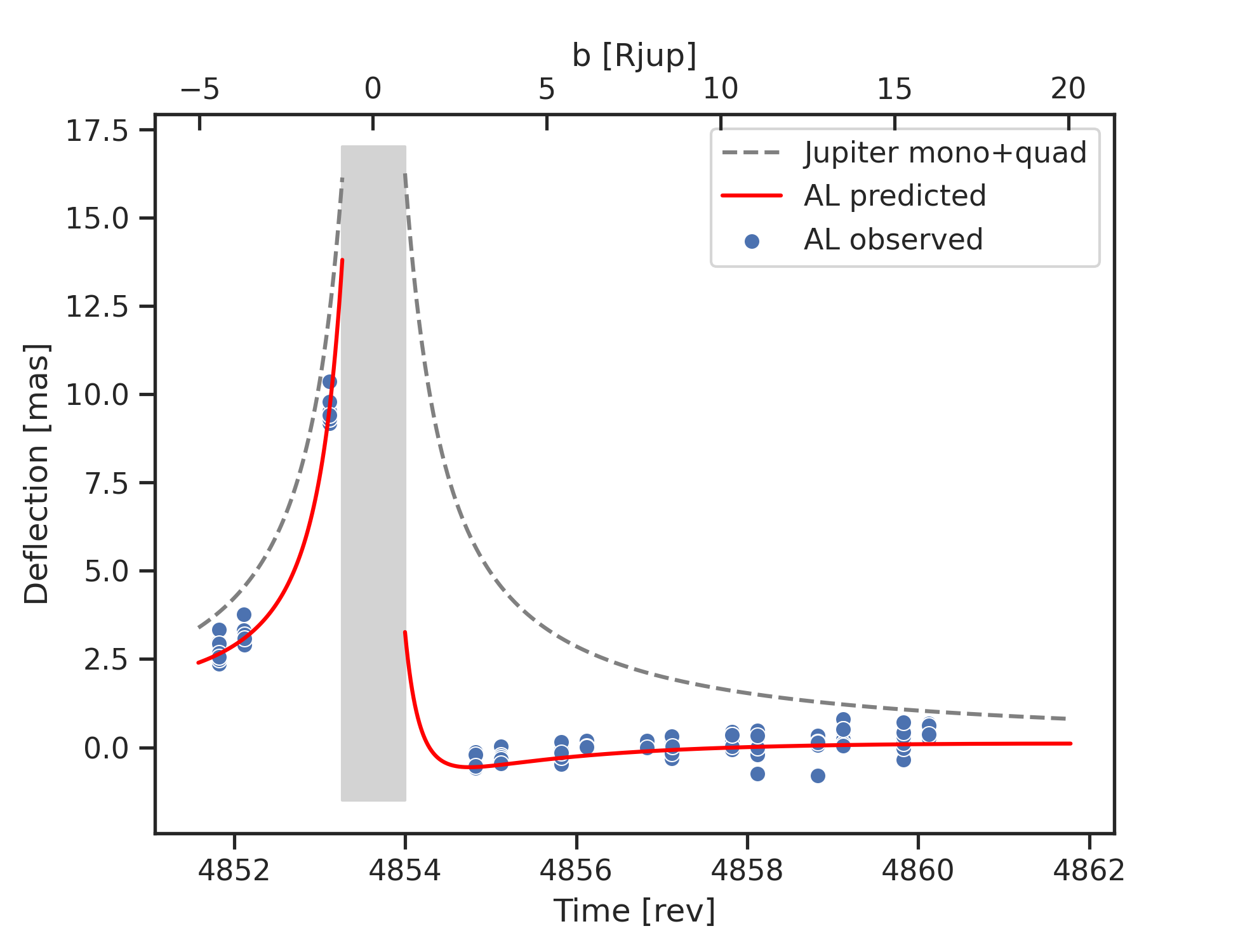} 
      \caption{The AL coordinate for the {\it target star}, $\eta_{00}'$, at the various observing times 
	plate-transformed to the reference frame calculated using 
	the best-fit linear plate parameters per transit.
	The blue circles depict the result obtained for the observations with the gray shaded area 
	showing the exclusion zone of Jupiter's disk.      
	 The gray dashed line 
	shows the predicted total light deflection due to Jupiter's monopole and quadrupole. The 
	red continuous line shows the combined deflection due to Jupiter and the Sun
	projected in the AL scan direction. The tick marks at the
	top show the impact parameter in units of the radius of Jupiter.}
  \label{fig:gareq_event}
\end{figure*}

\begin{figure}[h]
  \centering
      \includegraphics[scale=0.55]{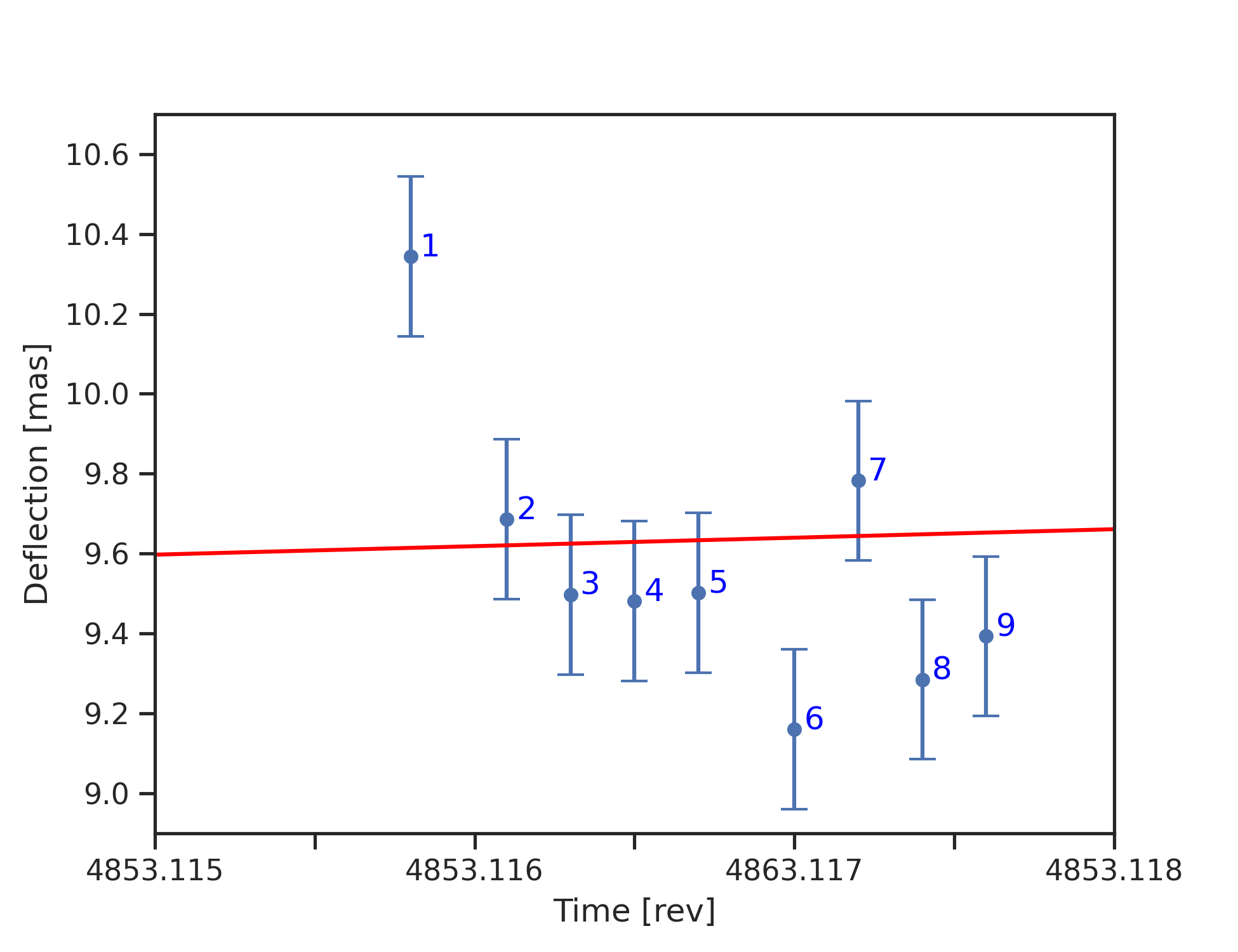} 
      \caption{A zoom-in on the highest deflection observed at the third transit shown in 
	Fig.\ref{fig:gareq_event} with the observations depicted 
	as blue circles and the predicted AL-light deflection by Jupiter as the red line.
	The numbers accompanying each blue circle denote the corresponding AF column associated with the 
	observation, e.g. 1 means that the measurement is associated with AF1.
	It can be seen that the AF1 measurement is an outlier at more than 3-$\sigma$.
	}
  \label{fig:gareq_event_zoomed}
\end{figure}

The set of 31 reference stars shown in Fig.~\ref{fig:1} are not always seen by Gaia whenever the {\it target star} is
seen. This is due to a variety of reasons; the way Gaia scans the sky, stars at the edge of the FOV,
observations that fall in the gap between two CCD columns etc.
Furthermore, stars seen on any given single transit do not necessarily have all the sequence of
9 AF observations for the chosen time interval around the {\it target star}.
This results in an inhomogeneous set of observations per frame, 
with some having barely a dozen reference stars in total.
We chose $\pm$40 secs as the time interval (obtained with the same CCD column) 
for reference stars surrounding the {\it target star} as
a good compromise between maximizing the number of reference stars and minimizing the higher order effects 
that come into play from larger fields.

Fig.~\ref{fig:gareq_event} depicts the transformed \footnote{The field angles are transformed onto the 
reference frame chosen to be the fourth transit. This choice is mainly dictated by the frame that has the maximum 
number of reference stars.} AL field angle, $\eta_{00}'$, for the {\it target star}
at the various observing times for the 15 transits in Table~\ref{table:1} using the
best-fit linear plate parameters obtained with the surrounding reference stars. 
For comparison, we show the overlapping prediction (red curve) obtained with the Gaia relativity model 
(GREM; \citealt{klioner2003, klioner2004})
which refers to the classical relativistic light deflection effect due to Jupiter and the Sun, with a dominant
contribution from Jupiter's monopole.
Also shown is the total light deflection (gray curve) due to Jupiter demonstrating the 
16.2 mas deflection that would be seen by a grazing light ray at the limb of Jupiter.
There is a good agreement between the observed and predicted AL deflections
for all the transits within the AF1-9 spreaded observations per transit.
We find that the signal at closest approach is a factor of 50 times that of the 0.2mas standard deviation of the 
observed values over AF2-9 at highest deflection (AF1 is removed as it is a clear outlier).
This is an unprecedented measurement at optical wavelengths in the 
literature of the relativistic deflection of light due to Jupiter, and the first time at any wavelength 
for a star $\sim$7" from its limb.

A zoomed-in view on the highest set of deflection observations seen on the third transit is shown in 
Fig.~\ref{fig:gareq_event_zoomed} clearly demonstrating that the AF1 measurement is an outlier at the 3-$\sigma$ level.
Once that point is removed the standard deviation of the points at highest deflection (0.2mas) is only slightly 
lower than that from the points in the remaining 14 transits ($\sim$0.25mas) and from the points in the 
last 12 transits ($\sim$0.22mas) that are minimally affected by Jupiter.

\newpage
\section{The reference stars residual distribution and error budget analysis}\label{sect:error_budget}

\begin{figure}
  \includegraphics[width=9cm]
                  {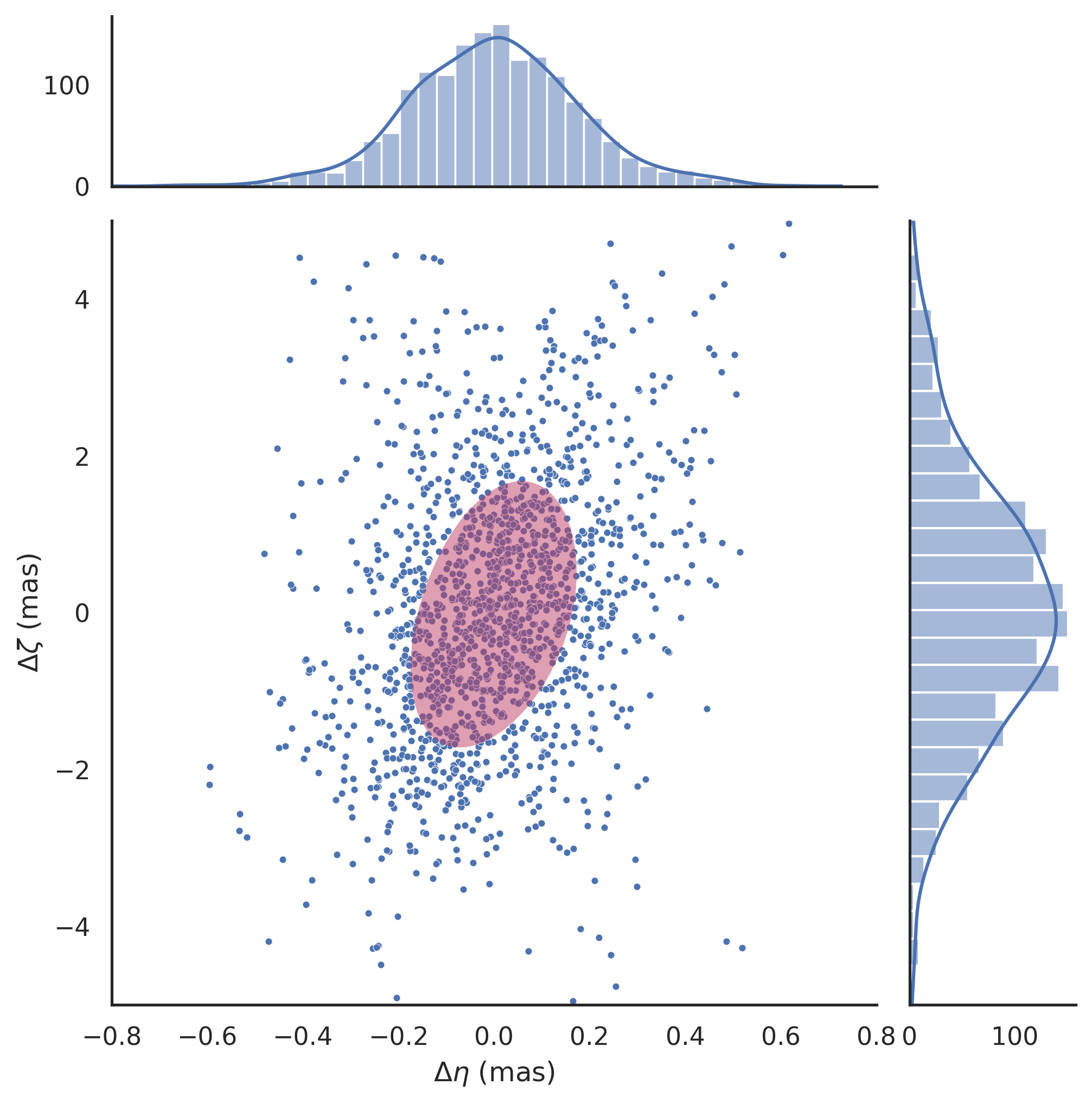}
      \caption{The joint AL-AC post-fit residual distribution of the reference stars in the first 15 transits 
	of the GAREQ field (further described in the main text) surrounding the {\it target star} and showing 
	$\sigma_{AL}$=180$\mu$as and $\sigma_{AC}$=1.822mas. 
	The red shaded ellipse shows the 1$\sigma$ joint uncertainty.  
	The histogram distributions are shown in the margins: AL at the top and AC on the right.}
  \label{fig:resids_event}
\end{figure}

\begin{figure}
  \includegraphics[width=9cm]
                  {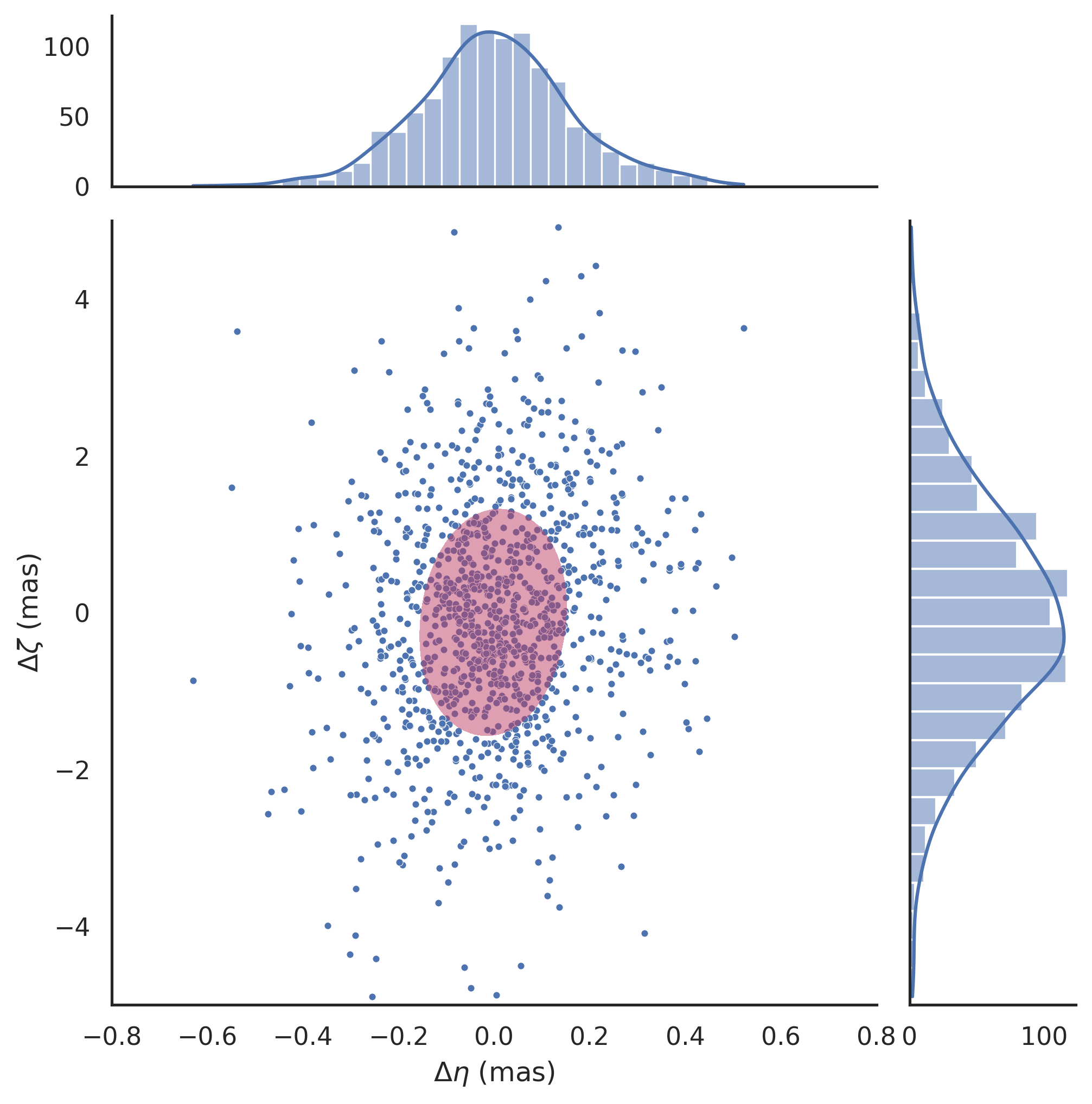}
      \caption{The joint AL-AC post-fit residual distribution of the reference stars in the last 10 transits
	of the GAREQ field (further described in the main text) surrounding the {\it target star} and 
	showing $\sigma_{AL}$=162$\mu$as and 
	$\sigma_{AC}$=1.469mas that is depicted as the red shaded ellipse in the form of the 1-$\sigma$ joint 
	uncertainty.
	The corresponding AL and AC histogram residual distributions are shown in the margins.}
  \label{fig:resids_10transits}
\end{figure}

The deflection signal seen in the {\it target star} depends mainly on the reference stars and their residual distribution,
the smaller the uncertainties the more tight are the estimated plate parameters that lead to a cleaner deflection 
signal.
The {\it target star} was seen on a total of 25 (mostly consecutive) transits on Feb 22-Feb 26 2017 thanks to the optimized scanning law (the first 15 of which are detailed in Table~\ref{table:1} and shown in Fig.~\ref{fig:gareq_event}).
We looked at the residual distribution of the reference stars surrounding the {\it target star} under two different circumstances: in the first 15 transits (that includes the highest deflection signal) that is shown in 
Fig.~\ref{fig:resids_event}, and, in the following 10 consecutive transits shown in 
Fig.~\ref{fig:resids_10transits}.

In the first 15 transits we find $\sigma_{AL}$=180$\mu$as and $\sigma_{AC}$=1.822mas in the post-fit residual distribution
with a Pearson's correlation between the AL and AC residual distributions of 0.275.
In the following 10 transits the values decrease to $\sigma_{AL}$=162$\mu$as and $\sigma_{AC}$=1.469mas
with a Pearson's correlation of 0.15 between the two residual distributions. Indeed, the bright disk of Jupiter causes extra systematic
effects that shows up as a very mild correlation in the AL and AC residual distributions of the surrounding reference stars, besides increasing the standard deviation of such a distribution. 
This extra systematic effect is expected to be accounted for or at least significantly reduced during regular cyclic developments in future data processing.
It can be seen that the scale of the AC axis is roughly 5 times larger than that of the AL axis. We have deliberately maintained this scale difference in the figures in order to better demonstrate the distribution and correlation of 
the residuals.

\begin{figure*}[htb]
\centering
    \subfigure[][]{
      \includegraphics[width=9cm]
                  {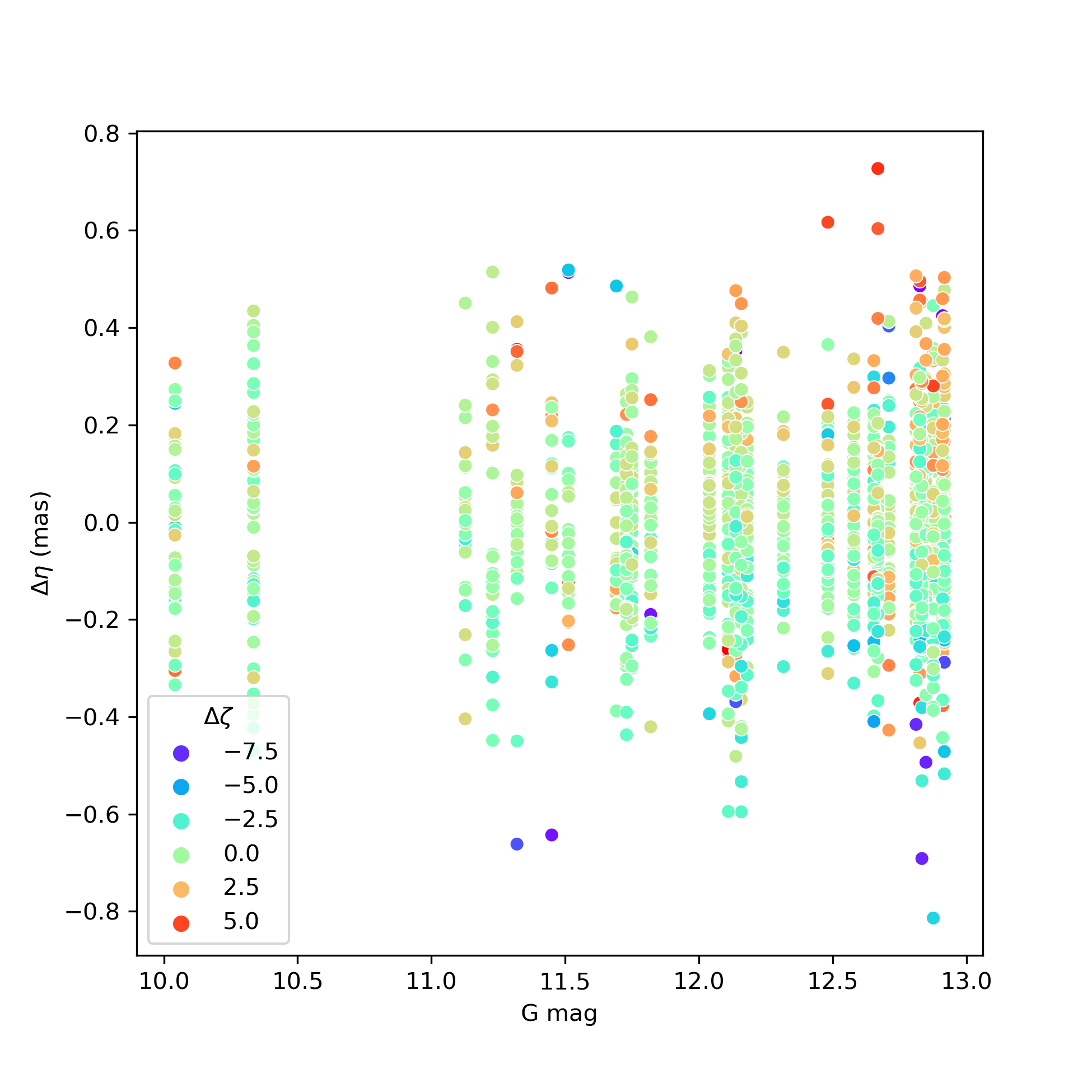} } 
     \subfigure[][]{
      \includegraphics[width=9cm]
                  {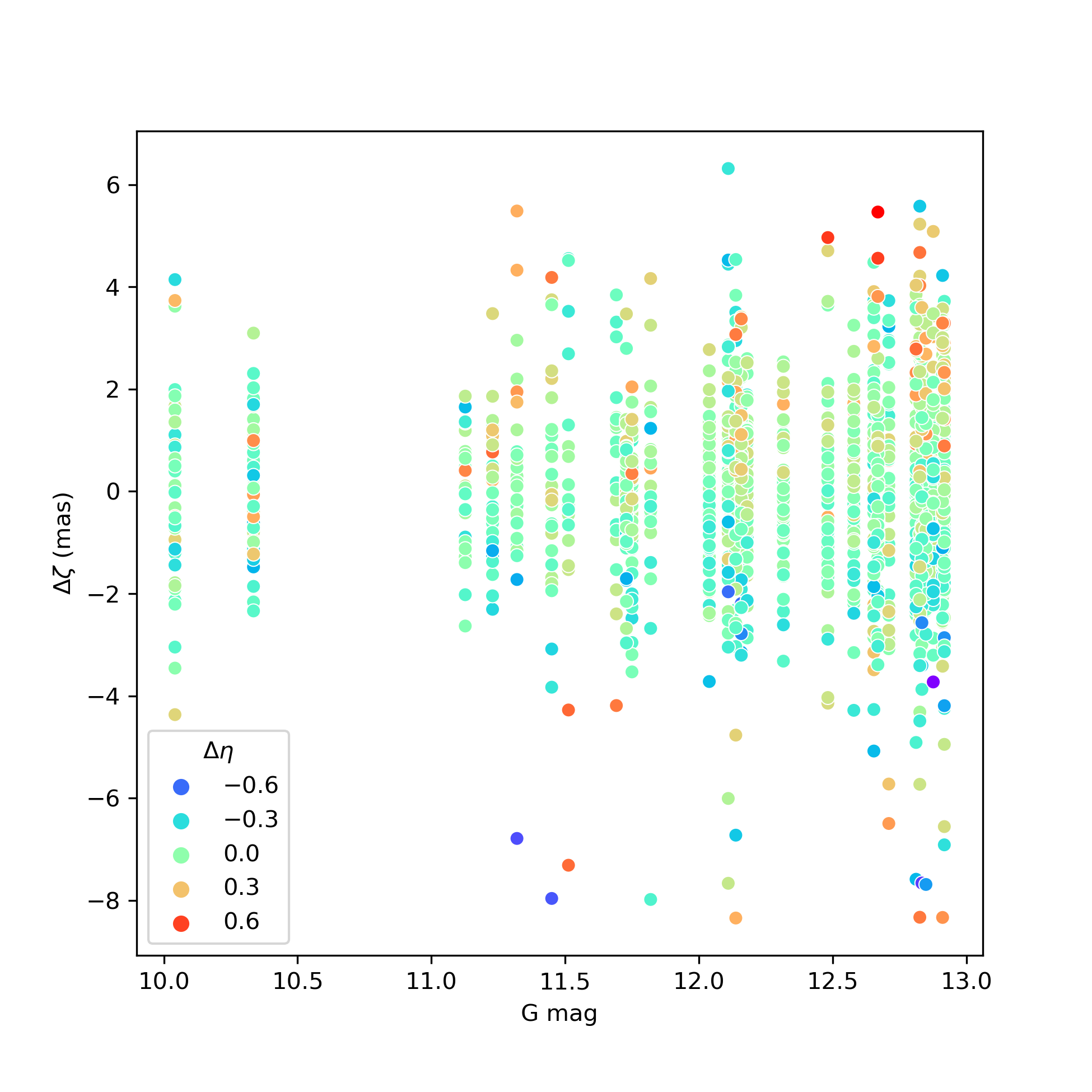} } 
     \caption{The post-fit AL (panel (a)) and AC (panel (b)) residuals of the reference stars in the 
	15 transits versus the magnitude of the star. The color coding is as a function of the AL/AC residual
	value.}
   \label{fig:detaVSgmag}
\end{figure*}

\subsection{Residual distributions as a function of magnitude}

The post-fit residuals in the AL and the AC scanning directions is shown as a function of the reference 
star's magnitude in Fig.~\ref{fig:detaVSgmag}. 
It can be seen that in general the outliers in the AL residual distribution have a darker shade of red
or blue indicating that they are also outliers in their AC residuals. 
The same is not necessarily true in the case of the AC residuals, i.e. the outliers do not always have
high values of $\Delta\eta$.

\subsection{Reference stars in a 'quiet' and normal field}
\begin{figure}
  \includegraphics[width=9cm]
                 {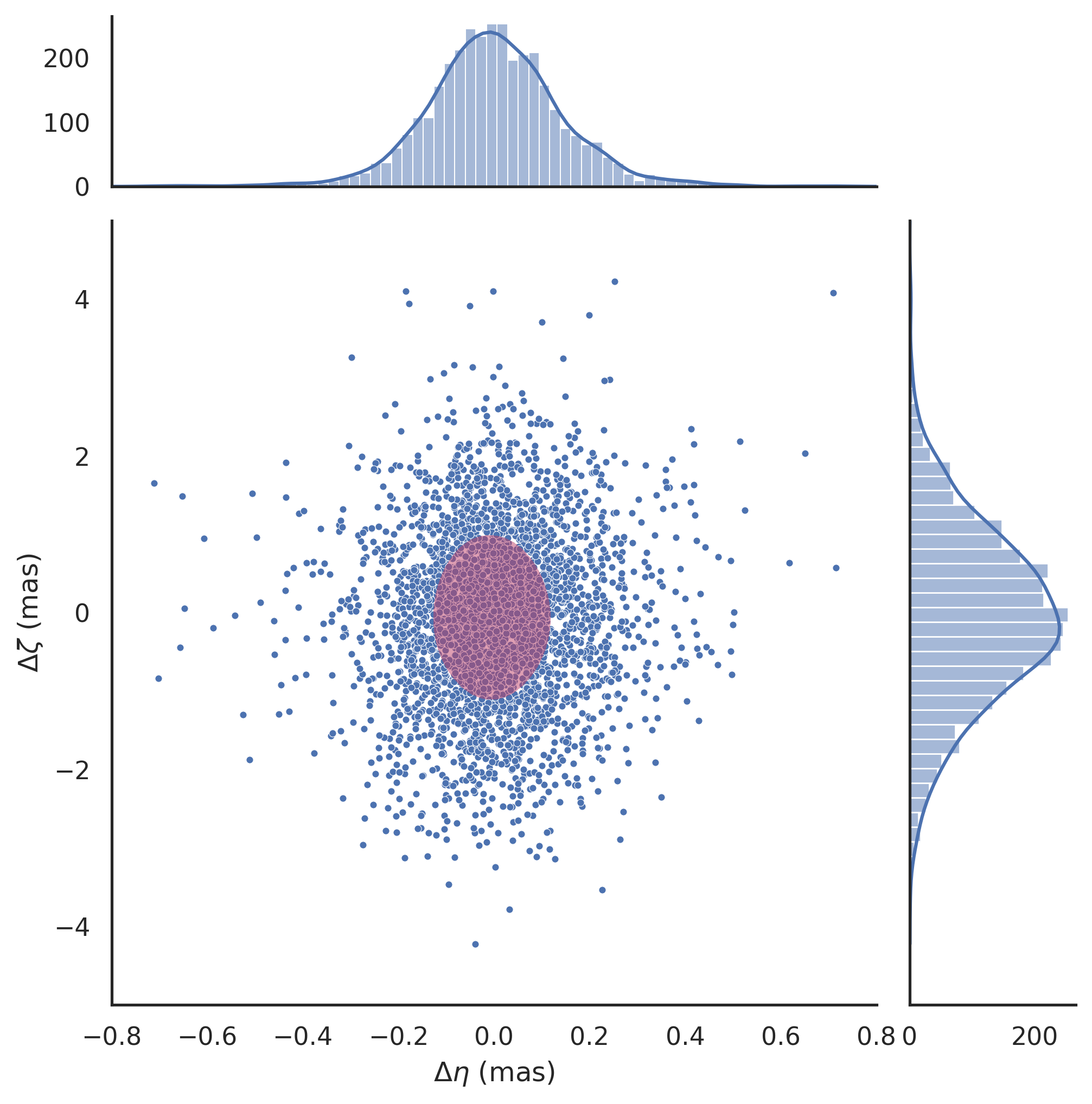}
      \caption{The joint AL-AC post-fit residual distribution of stars in a 'quiet' field showing 
	$\sigma_{AL}$=150$\mu$as and $\sigma_{AC}$=1.213mas depicted as the red shaded ellipse in the form
	of the 1-$\sigma$ joint uncertainty. 
	The corresponding histogram distributions are shown in the margins.}
  \label{fig:resids_nullfield}
\end{figure}

We looked at a set of stars in the same magnitude range (10 < G mag < 13), with a similar mean
magnitude ($\bar G\sim$ 12), in a different and 
`quiet’ field that was also seen on four consecutive transits. 
This is a field surrounding a star at RA: 92.95 DEC: 22.83 degs and is quite similar to the GAREQ field as it satisfies 
three main requirements: similar mean magnitude, high-cadence observations and a large number of (in this field about 200) bright reference stars. 
A large number is particularly important as not all reference stars are always seen on successive transits. 
For e.g. the last 2 transits have almost 180 common reference stars, whereas those in common with the 
first 2 transits are roughly 60. 
The joint residual distribution (considering goodness-of-fit values of the image fit to the 
PSF/LSF model of less than 100) is shown in Fig.~\ref{fig:resids_nullfield}.
Indeed, for such a clean field of stars with no extra disturbances the Pearsons correlation between the AL and 
AC residual distribution is 0.03 demonstrating that the presence of Jupiter within $\sim$500 arcseconds of the reference stars is enough to introduce extra systematic effects of the order of several tens of $\mu$as.

We performed a test by using a random subsample of 30 stars for this field matching the number of 
reference stars in the GAREQ field. This led to similar residuals (i.e. $\sigma_{AL}\sim$ 142$\mu$as) demonstrating 
that unmodeled systematics, probably induced by imperfect instrumental calibrations, are the dominant 
effect 
in line with the robust estimates of the actual standard deviations of the post-fit residuals in EDR3 (see Fig.~A.1 in 
\citealt{lindegren2021}).

\newpage
\section{Conclusions}\label{sect:conclusions}

In this paper we have developed a sophisticated differential astrometric model that 
is appropriate for a scanning space satellite such as Gaia. 
In general, the differential set-up allows to study single events in a small field,  
with a sufficient number of bright reference stars, such as the ones presented here.

This method relies upon a transformation of the astrometric observations of stars (in the form of calibrated field angles)
onto tangent planes at various times finally anchored to a common reference frame.
Only the commanded/nominal attitude of the spacecraft is used along with generalized least-squares 
in order to obtain the best-fit plate parameters and the floor of the reference stars. 
This establishes a small-field reference frame where any additional "systematic effect" can be
further studied.
We have used this framework to provide a preliminary assessment of the Gaia Relativistic 
Experiment on Quadrupole light deflection (GAREQ) event observed in
February 2017 by investigating:
a) the observability of the relativistic deflection of light close to Jupiter, and, 
b) Gaia’s astrometric capabilities under critical conditions such as those around an extremely bright extended object.

We use astrometric measurements provided by Gaia of 31 bright reference stars (10 $<$ G mag $<$ 13) over
$\sim$3 days all 
lying within a field of 0.8x1.3 degs surrounding a {\it target star} ($G=12.78$ mag) of particular interest. 
The bright magnitude limit for 
the reference stars was imposed mainly by the availability of 2d CCD cut-outs (windows in the Gaia jargon) providing precise 
2d observations, and secondarily by the overall random astrometric standard errors.
We use observations of the bright {\it target star} and surrounding reference stars  
in transits leading up to the time of closest approach to Jupiter's limb and on subsequent transits during the GAREQ event.
The best-fit plate parameters obtained using the model ultimately provide the {\it target star's} deflection as a systematic effect.

The {\it target star} was successfully observed several times within 5 Jupiter
radii from Jupiter’s estimated center, 0.35 Jupiter radii (roughly 6.7”) from Jupiter’s limb at closest
approach, and especially favorable for studying the effects on the relativistic deflection of light from the 
star as it passes near a giant oblate planet. 
These special events are being studied within the context of GAREQ where the ultimate goal is to disentangle a
$\sim$100$\mu$as AL
signal at closest approach to Jupiter's limb due to the planet's quadrupole moment
from the much larger $\sim$10mas monopole deflection. 

Here we present the first results of the relativistic light deflection at the $\sim$10mas level with 
a reference star residuals distribution of $\sigma_{AL}$ $\sim$180$\mu$as.
We are mainly dominated by Jupiter's monopole deflection signal with the present differential 
astrometric model using current DPAC processed data from cycle 3 (i.e. EDR3/DR3 astrometry).
The event of February 2017 presented here is the first of three predicted events expected to take place over 3 years of 
the Gaia nominal mission (the other two were slated for September 2018 and April 2019 
to be covered by future cyclic processed data, cycle 3 processed data only covers
until May 2017). 
Moreover, using data from the cyclic reprocessing in future
cycles, that will better handle the extra systematic effects arising from Jupiter's glare in the
field, we hope to better study Jupiter's quadrupole deflection. 

Our results show that the scene chosen for the GAREQ event  
allows for a clean detection of the relativistic deflection of light 
from a star by Jupiter with a S/N of $\sim$50 and in good agreement 
with the predicted total deflection. 
This detection is unprecedented as it is: a) the closest ever to Jupiter’s limb ($\sim$7") in the optical, and,
b) with the highest S/N at any wavelength. 
These results strongly encourage further investigations
towards searching for and disentangling the quadrupole deflection from the total signal
using Gaia astrometric observations from future processing cycles.

\begin{acknowledgements}
We thank the referee for useful suggestions that helped to improve the paper.
We further thank M. Biermann and U. Bastian for valuable suggestions and discussions that helped us 
refine the differential methodology and the AGIS and IDU teams for the processing and calibration of the data
used in this paper.
We also thank F. Mignard and S. Klioner for their essential role in the GAREQ experiment and for the 
optimization of the Gaia scanning law that led to the deflection event of February 2017, and C. Crowley 
and his team for demonstrating the exceptional performance of Gaia under the exteme observing circumstances
of the scene around Jupiter that led to the successful repeated observations of the target star, 
including the measurement at
0.35 $R_{Jupiter}$ away from the planet's limb. 
We acknowledge the support of A. Brown, T. Prusti and G. Gracia throughout the development of this work.
UA thanks Alessandro Sozzetti for helpful discussions and brainstorming sessions that 
helped with an effective implementation of crucial stages of data treatment.

This work was financially supported by the European Space Agency (ESA) in the framework of the Gaia project;
the Italian Space Agency (ASI) through Gaia mission contracts: 
the Italian participation to the Gaia Data Processing and Analysis Consortium (DPAC), 
ASI 2014-025-R.1.2015 and ASI 2018-24-HH.0 in collaboration with the Italian National Institute of Astrophysics.

This work has made use of data from the European Space Agency (ESA) mission Gaia (https://www.cosmos.esa.int/gaia), 
processed by the Gaia Data Processing and Analysis Consortium (DPAC, https:
//www.cosmos.esa.int/web/gaia/dpac/consortium). 
Funding for the DPAC has been provided by national institutions, in particular the 
institutions participating in the Gaia Multilateral Agreement. 

Python libraries used: Matplotlib \citep{hunter2007}, NumPy \citep{harris2020},
Pandas \citep{reback2021} and Seaborn \citep{waskom2021}.
\end{acknowledgements}

%
%

\newpage

\bibliographystyle{aa}
\bibliography{gareq}

\end{document}